\documentclass[%
 reprint,
superscriptaddress,
nofootinbib,
 amsmath,amssymb,
 aps,
prd,
]{revtex4-2}
\usepackage{color}
\usepackage{graphicx}
\usepackage{graphics}
\usepackage{caption}
\usepackage{subcaption}
\usepackage{dcolumn}
\usepackage{bm}
\RequirePackage{mathptmx}      
\RequirePackage{amsmath,amssymb}
\RequirePackage{bm}
\RequirePackage{bbm}
\RequirePackage{flushend}
\usepackage{multirow}
\RequirePackage{natbib}
\usepackage[colorlinks,citecolor=blue,urlcolor=blue,linkcolor=blue]{hyperref}
\RequirePackage[usenames, dvipsnames]{xcolor}
\RequirePackage{tikz}
\RequirePackage[percent]{overpic}
\RequirePackage{setspace}
\usetikzlibrary{shapes.geometric, arrows}
\usepackage{natbib}
\usepackage{bigints}
\usepackage{mathrsfs}
\usepackage{xr}
\usepackage[colorlinks,citecolor=blue,urlcolor=blue,linkcolor=blue]{hyperref}
\usepackage{amsfonts}

\newcommand{\vQn}{v_{Q,{n}}}

\newcommand{\Chi}{\mathrm{X}}
\newcommand{\Real}{\mathbb{R}}

\begin{document}

\widetext
\leftline{Version xx as of \today}
\leftline{Author: Sara Algeri}

\title{K-2 rotated goodness-of-fit for multivariate data}

\author{Sara Algeri}

\affiliation{School of Statistics, University of Minnesota, Minneapolis (MN), 55455, USA. Email: salgeri@umn.edu }
\date{\today}

\begin{abstract}
Consider a set  of multivariate distributions, $F_1,\dots,F_M$, aiming to explain the same phenomenon. For instance, each $F_m$ may correspond to a  different candidate background model for calibration data, or to one of  many possible signal models we aim to validate on experimental data. 
In this article, we show  that tests for a wide class of apparently different models $F_{m}$  can be mapped into a single test for a reference distribution $Q$. As a result,   valid inference for each $F_m$  can be obtained by simulating \underline{only} the distribution of the test statistic under $Q$. Furthermore, $Q$ can be chosen conveniently simple to substantially reduce the computational time. 
\end{abstract}


\maketitle

\section{Introduction}
\label{intro}
 Despite the popularity of classical goodness-of-fit tests such as Pearson's $\Chi^2$ \cite{pearson}, likelihood ratio and Kolmogorov-Smirnov \cite{kolmogorov,smirnov}, their applicability often face serious challenges in many situations relevant to modern experiments. For instance, when conducting multidimensional searches  in a binned data regime, the limited sample size may affect the validity of the $\chi^2$ approximation for $\Chi^2$. Moreover, if the expected number of events is small,   the $\Chi^2$ statistics may be biased, that is, its power can be smaller than the prescribed significance level \cite{haberman}. Unfortunately, this may occur even when a reasonable $\chi^2$ approximation for it exists, leaving little hope when aiming to address the problem by means of Monte Carlo simulations.   Similarly, the likelihood ratio may suffer from additional biases due to the estimation of the unknown parameters \cite[e.g.,][]{cressie}.  These problems can often be overcome in the unbinned data regime by means of tests such as Kolmogorov-Smirnov, Cramer-von-Mises, and Anderson-Darling. In this case, the price to pay is the  loss of distribution-freeness when the models under study are multivariate and/or involve unkown parameters that need to be estimated. As a result, one needs  to derive or simulate the distribution of the  test statistic on  a case-by-case basis.

In this article, we discuss a simulation-based testing strategy which allows us to overcome all these short-comings and equips experimentalists with a novel  tool to perform goodness-of-fit  while reducing substantially the computational costs.  
 The rationale behind the solution  is somewhat close in spirit (but different in nature) to that of the well-known Metropolis-Hasting algorithm \cite{metropolis,hastings}. When aiming to sample data from a complex distribution $F$, the  Metropolis-Hasting algorithm circumvents the difficulties associated with sampling directly from $F$  by considering a much simpler distribution $Q$. The choice of $Q$ is  arbitrary and thus one can often compute integrals in $F$, or approximate the latter solely relying on samples from $Q$.   In a similar manner,  the tests presented here consist of converting the testing problem for a given distribution $F$ into a test for a \emph{reference-distribution} $Q$. We show that tests for many different distributions $F_1,\dots,F_M$  can all be mapped into one single test for $Q$. Also in this case, $Q$ can be chosen conveniently simple. It follows that one can calculate the prescribed test statistic on the  data  for one or more candidate models $F_m$,  and compare its observed value directly with the simulated distribution of the test statistic under $Q$, avoiding $M$ separate simulations.

From a theoretical stand-point, the key element of the solution is the \emph{Khmaladze-2} (K-2) \emph{transform}\footnote{The Khmaladze-2 transformation has not to be confused with the well-known ``Khmaladze transformation'', also referred to in literature as Khmaladze-1 (K-1) transform, and originally proposed by the same author in \cite{khm82}. }, also known as \emph{Khmaladze's rotation},  a novel unitary-transformation for empirical processes introduced in recent years by   \cite{khm16,khm17}.  
The test statistics proposed in this article are  extensions of the Kolmogorov, Cramer-von-Mises and Anderson-Darling's statistics and adequately constructed to  account for the variability associated with the estimation of the parameters. For the specific case of Anderson-Darling, we will see that the reference distribution $Q$ also plays the role of weighting function. That is,  it can be used to assign the desired weights to the tails of the distribution. Finally, we evaluate the performance of the tests proposed through a suite of simulation studies.

The remainder of the manuscript is organized as follows. In Section \ref{standard} we provide an overview on the classical empirical process, that is, the main object at the core of classical goodness-of-fit tests. Section \ref{multi} is devoted to extend the classical empirical process to the multivariate parametric setting and introduces the projected empirical process. While the latter is shown to provide remarkable computational advantages, its main relevance for us is that of setting the ground to perform distribution-free goodness-of-fit. Distribution-freeness is the focus of Section \ref{FtoQ}. There, we introduce the K-2 transform and investigate its properties through a suite of simulation studies. Some final remarks are collected in Section \ref{Final}. Details on  mathematical derivations  are provided in the Appendix. 

\section{The classical empirical process}
\label{standard}
 Consider a   sample $x_1,\dots,x_n$ for which each measurement $x_i$ is the realization of a random variable $X_i$. For the moment, we assume that the $X_i$s take values on the interval $[L,U]$, are independent and identically distributed (i.i.d.) with  cumulative distribution function (cdf), $P$,  either continuous or discrete. In this setup, the empirical process is
\begin{equation}
\label{vn}
v_{P,n}(x)=\sqrt{n}[P_n(x)-P(x)]=\frac{1}{\sqrt{n}}\sum_{i=1}^n\bigl[\mathbbm{1}_{\{x_i\leq x\}}-P(x)\bigl]
\end{equation}
 where $P_n(x)=\frac{1}{n}\sum_{i=1}^n\mathbbm{1}_{\{x_i\leq x\}}$ is the empirical cumularive distribution of $x_1,\dots,x_n$ and which is known to converge to $P$, when $n\rightarrow \infty$. From the first equality in \eqref{vn}, it is clear  that, for every point in $[L,U]$, $v_n$ consists of a  ``magnified'' difference between the empirical cumulative distribution of the data and  $P$, where the ``magnifying factor'' is $\sqrt{n}$. Hence, when replacing $P$ with any  $F\not\equiv P$, the differences between $P_n$ and $F$ becomes more and more obvious as $n\rightarrow\infty$.
 
 The literature investigating the properties  of $v_n$ is vast (see \citet{wellnerrev} for a review), and mainly focuses on the case where $F$ is fixed. 
In practical applications, however, $F$ typically   depends on unknown parameters to be estimated. It is therefore important to extend \eqref{vn} to this setting.

\section{The multivariate parametric regime}
\label{multi}  
Consider a sample of i.i.d. observations over the search region $\mathcal{\bm{X}}\subseteq \Real^d$ and let $P(\bm{x})=P(x_1,\dots,x_d)$ be their true underlying distribution. Despite $P$ is unknown, suppose we are given a simplified candidate model $Q_{\bm{\theta}}(\bm{x})$  for the data, with $\bm{\theta}$ being a set of $p$ unknown parameters,  and let $q_{\bm{\theta}}(\bm{x})$ be the respective probability density function (pdf)   or probability mass function (pmf). We assume that $Q_{\bm{\theta}}$ is easy to simulate from, to evaluate, and to estimate its parameters. 
For instance, $Q_{\bm{\theta}}$ may  be the cdf of a $d$-dimensional normal distribution with independent components, known variance and mean vector depending on $\bm{\theta}$. 
Moreover, suppose another model, $F_{\bm{\beta}}$, is given and let $\bm{\beta}$ be the set of parameters charachterizing it. The distribution $F_{\bm{\beta}}$ may be arbitrarily complex and, potentially, much harder to simulate from, to estimate, and even to evaluate than $Q_{\bm{\theta}}$. In this section and those to follow, we will show that we can construct two test statistics, one to test $F_{\bm{\beta}}$ and one to test $Q_{\bm{\theta}}$, whose  null distribution is the same. 
In order to achieve this goal we begin by constructing a  test for $Q_{\bm{\theta}}$ based on the so-called \emph{projected empirical process}. 
\subsection{The projected empirical process}
An extension of \eqref{vn} to this setup is  given by the parametric empirical process
\begin{align}
\label{vQn}
v_{Q,n}(\bm{x},\bm{\theta})&=\frac{1}{\sqrt{n}}\sum_{i=1}^n\psi_{\bm{x},\bm{\theta}}(\bm{x}_i)\quad\text{with}\\
\label{phi}
\psi_{\bm{x},\bm{\theta}}(\bm{x}_i)&=\bigl[\mathbbm{1}_{\{\bm{x}_i\leq \bm{x}\}}-Q_{\bm{\theta}}(\bm{x})\bigl]
\end{align}
and  $\mathbbm{1}_{\{\bm{x}_i\leq \bm{x}\}}=\mathbbm{1}_{\{x_{1i}\leq x_1,\dots,x_{pi}\leq x_p\}}$ takes value one for all the data points whose  coordinates are smaller or equal  than $\bm{x}=(x_1,\dots,x_d)$, and zero otherwise. 

Denote with  $\widehat{\bm{\theta}}$ be the maximum likelihood estimate (MLE) of  ${\bm{\theta}}$, which we assume satisfies the classical regularity conditions  \cite[e.g.,][p. 500]{cramer} (see also \cite{nature} for a high-level review). We denote the score vector of $Q_{\bm{\theta}}$ with $\bm{u}_{\bm{\theta}}$, i.e.,
\begin{align}
\label{scoreb2}
\bm{u}_{\bm{\theta}}(\bm{x})&=\bigl[u_{\bm{\theta}_1}(\bm{x}),\dots,u_{\bm{\theta}_p}(\bm{x})\bigl]^{T}
\end{align}
where each element $u_{\bm{\theta}_j}(\bm{x})$ corresponds to
\begin{equation}
\label{components}
u_{\bm{\theta}_j}(\bm{x})=\frac{\partial }{\partial \theta_j}\log q_{\bm{\theta}}(\bm{x})
\end{equation}
with  $\theta_j$, $j=1,\dots,p$ being the components of the parameter vector $\bm{\theta}$. We denote with
 $\Gamma_{\bm{\theta}}$  the Fisher-information matrix, i.e., the matrix of elements
\begin{align}
\label{Gamma}
\Gamma_{\bm{\theta}{jk}}=\bigl\langle u_{\bm{\theta}_j},u_{\bm{\theta}_k}\bigl\rangle_{Q_{\bm{\theta}}}.
\end{align}
The inner product in \eqref{Gamma} is defined as
\begin{align}
\label{inner1}
\langle g,h \rangle_{Q_{\bm{\theta}}}&=\int_{\mathcal{\bm{X}}} g(\bm{t})h(\bm{t})q_{\bm{\theta}}(\bm{t})\text{d}\bm{t} \quad \text{if $Q_{\bm{\theta}}$ is continuous.}
\end{align}
 If $Q_{\bm{\theta}}$ is discrete, the integral in \eqref{inner1} is replaced by a summation over all the points of the search region $\mathcal{\bm{X}}$.
Lastly, we consider the normalized score function
\begin{align}
\label{scoreb}
\bm{b}_{\bm{\theta}}(\bm{x})&=\Gamma^{-1/2}_{\bm{\theta}}\bm{u}_{\bm{\theta}}(\bm{x})
\end{align}
and we denote with  $b_{\bm{\theta}_j}(\bm{x})$, $j=1,\dots,p$, its components.
The operation in equation \eqref{scoreb} consists of normalizing the vector $\bm{u}_{\bm{\theta}}$
in \eqref{scoreb2} by multiplying it by the inverse of the square root matrix of the Fisher information\footnote{In the applications to follow, the square root matrix has been computed via the Schur method \citep[e.g.,][Ch. 6]{higham}. Nonetheless, other methods to construct the square root matrix, such as diagonalization, Jordan decomposition, etc, are also viable options.}. The resulting functional vecor $\bm{b}_{\bm{\theta}}$ in \eqref{scoreb} consists of the normalized score functions $b_{\bm{\theta}_j}$, which have mean zero, unit variance, and are uncorrelated from one another under model $Q_{\bm{\theta}}$.

It was shown in  \cite{khm80} that, when replacing $\bm{\theta}$  in  \eqref{vQn} with $\widehat{\bm{\theta}}$, the resulting 
process, namely $v_{Q,n}(\bm{x},\widehat{\bm{\theta}})$, can be rewritten as  a projection of $v_{Q,n}(\bm{x},\bm{\theta})$ parallel to the normalized score  functions $b_{\bm{\theta}_j}$. Specifically,  a Taylor expansion and suitable algebraic manipulations lead to

\begin{align}
\label{projection0a}
v_{Q,n}(\bm{x},\widehat{\bm{\theta}})&\approx v_{Q,n}(\bm{x},\bm{\theta})-\frac{1}{\sqrt{n}}\sum_{j=1}^p\sum_{i=1}^nb_{\bm{\theta}_j}(\bm{x}_i){\langle b_{\bm{\theta}_j},\psi_{\bm{x},\bm{\theta}}\rangle}_{Q_{\bm{\theta}}},
\end{align}

\noindent where the error of the approximation is $o_p(1)$\footnote{The notation $o_p(1)$ is an abbreviation used in statistics to indicate that a sequence of
random vectors converges to zero in probability. In general, given two random sequences $R_n$ and $S_n$, we write $R_n=o_p(S_n)$ to indicate that $\frac{R_n}{S_n}$   converges  in probability to zero.}, that is, it quickly converges to zero in probability. The inner product in \eqref{projection0a} can be computed as in \eqref{inner1}. Details  on the derivation of \eqref{projection0a} are provided in Appendix \ref{app}.

It follows that, given the set of functions
\begin{equation}
\label{psihat}
\widetilde{\psi} _{\bm{x},\bm{\theta}}(\bm{t})=\psi_{\bm{x},{\bm{\theta}}}(\bm{t})-\sum_{j=1}^pb_{\bm{\theta}_j}(\bm{t}){\langle b_{\bm{\theta}_j},\psi_{\bm{x},\bm{\theta}}\rangle _{Q_{\bm{\theta}}}},
\end{equation}
we can specify the projected empirical process $\widetilde{v}_n(\bm{x},{\bm{\theta}})$ as
\begin{align}
\label{projection1a}
\widetilde{v}_{Q,n}(\bm{x},{\bm{\theta}})&=\frac{1}{\sqrt{n}}\sum_{i=1}^n\widetilde{\psi} _{\bm{x},\bm{\theta}}(\bm{x}_i),
\end{align}
and it is such that
\begin{align}
\label{projection1}
v_{Q,n}(\bm{x},\widehat{\bm{\theta}})&=\widetilde{v}_{Q,n}(\bm{x},{\bm{\theta}})+o_p(1);
\end{align}
hence, $v_{Q,n}(\bm{x},\widehat{\bm{\theta}})$ and $\widetilde{v}_{Q,n}(\bm{x},{\bm{\theta}})$ have the same asymptotic distribution.
\subsection{Testing $Q$}
\label{testingQ}
A notable advantage of working with empirical processes is that they allow us to construct an entire family of goodness-of-fit tests. 
For instance, to test the hypothesis $H_0:P=Q_{\bm{\theta}}$, many different test statistics can be constructed by simply taking functionals of $\widetilde{v}_{Q,n}(\bm{x},\bm{\theta})$. Some of these tests will be more powerful then others with respect to different alternatives, and thus, it is particulalry valuable to be able to access a variety of them.
Here, we focus on three main statistics which can be seen as a generalization of  Kolmogorov-Smirnov,  Cramer-von Mises, and Anderson-Darling's statistics, i.e.,
\begin{equation}
\begin{split}
\label{DQ}
\widehat{D}_Q&=\sup_{\bm{x}}|\widetilde{v}_{Q,n}(\bm{x},{\bm{\theta}})|,\quad \widehat{\omega}^2_Q=\int_{\mathcal{X}}\widetilde{v}^2_{Q,n}(\bm{x},{\bm{\theta}})q_{\bm{\theta}}(\bm{x})\text{d}\bm{x}, \\
\text{and}&\quad \widehat{A}^2_Q=\int_{\mathcal{X}}\widetilde{v}^2_{Q,n}(\bm{x},{\bm{\theta}})w_{\bm{\theta}}(\bm{x})q_{\bm{\theta}}(\bm{x})\text{d}\bm{x}
\end{split}
\end{equation}
with $w_{\bm{\theta}}(\bm{x})=[Q_{\bm{\theta}}(\bm{x})\bigl(1-Q_{\bm{\theta}}(\bm{x})\bigl)]^{-1}$ being the weighting function which allows us to highlight differences between the empirical cumulative distribution and $Q_{\bm{\theta}}$ in the tails.
 
  It is worth emphasizing that, in principle, one can use as test statistics the equivalent of those in \eqref{DQ} with $\widetilde{v}_{Q,n}(\bm{x},{\bm{\theta}})$ replaced by $\vQn(\bm{x},\widehat{\bm{\theta}})$. There are, however, two main advantages of working with $\widetilde{v}_{Q,n}(\bm{x},{\bm{\theta}})$ instead of $\vQn(\bm{x},\widehat{\bm{\theta}})$. First of all, as we will discuss in details in Section \ref{FtoQ}, $\widetilde{v}_n(\bm{x},{\bm{\theta}})$ sets the foundation to perform distribution-free tests. Second, $\widetilde{v}_{Q,n}(\bm{x},{\bm{\theta}})$ provides substantial gain, compared to $\vQn(\bm{x},\widehat{\bm{\theta}})$, from a computational stand point. 
 \begin{figure}[!h]
      \includegraphics[width=0.45\textwidth]{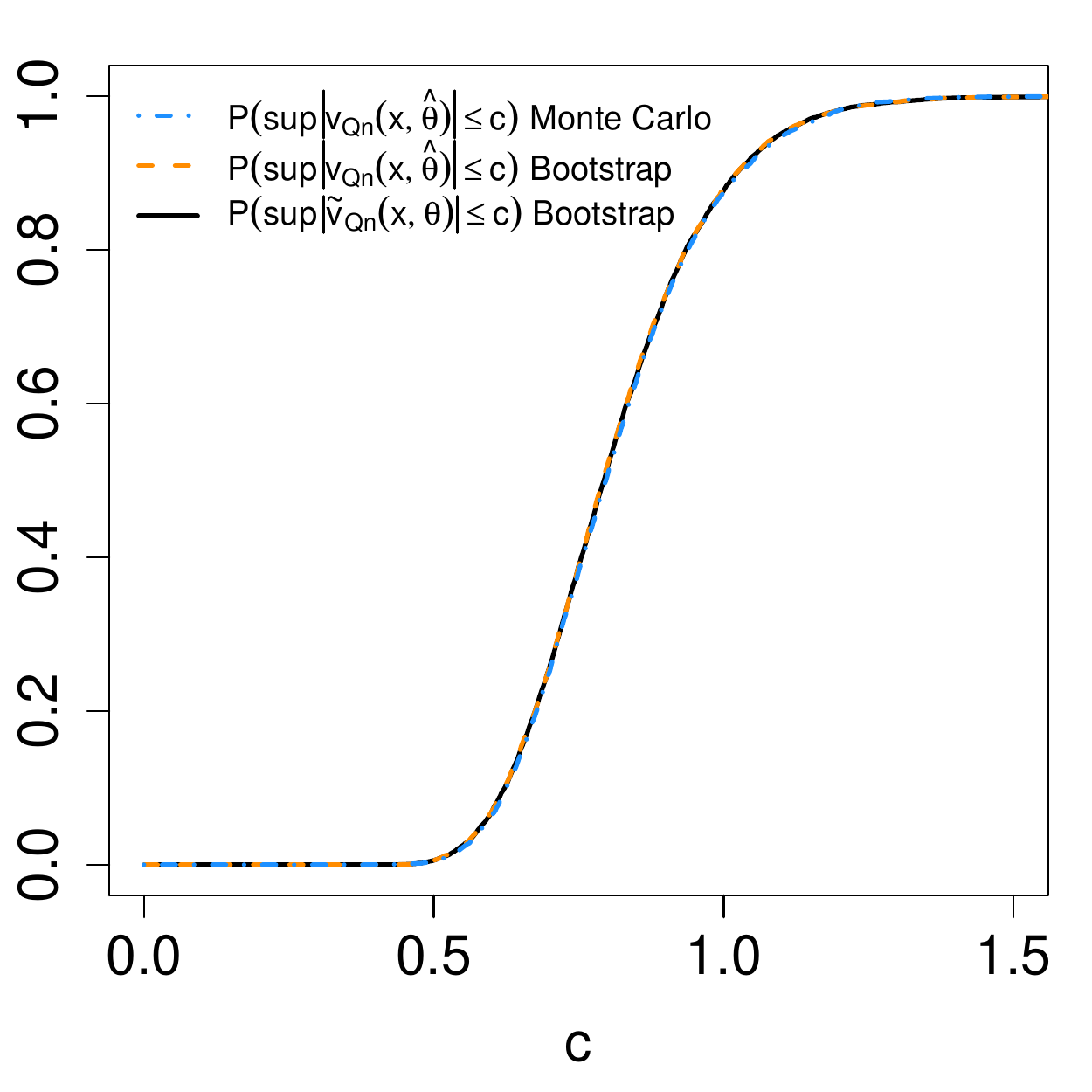}
      \caption{Comparing the boostraped distributions of the Kolmogorov-Smirnov's statistics $\sup_{\bm{x}}|\widetilde{v}_{Q,n}(\bm{x},{\bm{\theta}})|$ and $\sup_{\bm{x}}|\vQn(\bm{x},\widehat{\bm{\theta}})|$ and  the  distribution of $\sup_{\bm{x}}|\vQn(\bm{x},\widehat{\bm{\theta}})|$ simulated via Monte Carlo. In all three cases the simulation consist of $10,000$ replicates and the sample size is $n=100$.}
      \label{appQ}
\end{figure}

\begin{table}[!h]
\fontsize{9}{9}\selectfont
\begin{tabular}{!{\vrule width 1.5pt}c!{\vrule width 1.4pt}c|c!{\vrule width 1.5pt}}
                                 \noalign{\hrule height 1.5pt}
             &&\\[-1ex]
                  &$\sup_{\bm{x}}|\widetilde{v}_{Q,n}(\bm{x},{\bm{\theta}})|$&$\sup_{\bm{x}}|\vQn(\bm{x},\widehat{\bm{\theta}})|$\\[-1ex]
             &&\\[-1ex]
\hline
             &&\\[-1ex]
CPU time  &$9.429$ mins &$12.198$ hrs\\[-1.5ex]
                           &&\\
    
                                                    \noalign{\hrule height 1.4pt}
 \end{tabular}
\caption[Table 1]{Overall (system+user)  CPU time needed to simulate the distributions of the test statistics $\sup_{\bm{x}}|\widetilde{v}_{Q,n}(\bm{x},{\bm{\theta}})|$ and $\sup_{\bm{x}}|\vQn(\bm{x},\widehat{\bm{\theta}})|$ via the parametric bootstrap over $10,000$ replicates and   $n=100$ observations.}
\label{appQtab}
\end{table} 
Specifically,  in both cases, since $\bm{\theta}$ is unknown, one needs to simulate the distribution of the test statistics by means of the paramteric bootstrap, that is, we compute the MLE of $\bm{\theta}$ on the data observed, namely $\widehat{\bm{\theta}}_{\text{obs}}$, and, at each replicate, we sample  datasets from $Q_{\widehat{\bm{\theta}}_{\text{obs}}}(\bm{x})$.  The bootstrap procedure has been proven to lead to consistent results under very general conditions by \citet{babu}. They have shown that by simulating the distribution of continuous functionals of the parametric empirical process one can recover their true distribution if the parameters are estimated via MLE and the classical regularity conditions \cite[e.g.,][p. 500]{cramer} hold.

 When working with  $\vQn(\bm{x},\widehat{\bm{\theta}})$, to account for the variability introduced by the estimation process, one needs to repeat the maximization of the likelihood on each simulated bootstrap sample.
 Moreover, at each replicate, the cdf $Q_{\bm{\theta}}$ also needs to be evaluated on each point $\bm{x}\in \mathcal{X}$ considered, and  with $\bm{\theta}$ replaced by its estimated value on the simulated bootstrap sample. 
On the other hand, when working with $\widetilde{v}_{Q,n}(\bm{x},{\bm{\theta}})$, to account for the uncertainty associated with the estimation of $\bm{\theta}$,  instead of maximizing the likelihood at each iteration, we only need  to evaluate the normalized score functions in $\bm{b}_{\widehat{\bm{\theta}}_{\text{obs}}}(\bm{x})$ on each  simulated samples. Furthermore, despite we still  need to evaluate    $Q_{\bm{\theta}}$ at each $\bm{x}\in \mathcal{X}$ considered, as well as the integrals/summations in ${\langle b_{\bm{\theta}_j},\psi_{\bm{x},\bm{\theta}}\rangle _{Q_{\bm{\theta}}}}$, these only need to be computed  once, that is, for $\bm{\theta}=\bm{{\widehat{\bm{\theta}}_{\text{obs}}}}$,  reducing substantially the computational time. This approach is particularly advantageous since the error of approximating $\vQn(\bm{x},\widehat{\bm{\theta}})$ with $\widetilde{v}_{Q,n}(\bm{x},{\bm{\theta}})$  is only $o_p(1)$ (see equation \eqref{projection1}), and thus, it is negligible even for samples which are only moderately large.

To  illustrate these aspects with a toy example,   let $Q$ be the distribution of a  bivariate normal with independent components, truncated over the  region $\mathcal{X}=[1,20]\times[1,25]$, and with density
\begin{equation}
\label{q}
q_{\bm{\theta}}(\bm{x})\propto e^{-\frac{1}{2\theta_3}\bigl[(x_1-\theta_1)^2+(x_2-\theta_2)^2\bigl]},
\end{equation}
We draw a sample of $n=100$ observations from \eqref{q} with $\bm{\theta}=(-2,5, 25)$, and which will be considered our ``observed data''. We estimate $\bm{\theta}$ on such sample and we obtain $\widehat{\theta}_{obs}=(-0.77,  6.32, 22.02)$. 
We proceed by simulating the distribution of the Kolmogorov-Smirnov's statistics, $\sup_{\bm{x}}|\widetilde{v}_{Q,n}(\bm{x},{\bm{\theta}})|$ and $\sup_{\bm{x}}|\vQn(\bm{x},\widehat{\bm{\theta}})|$, via the parametric bootstrap. To emphasize the validity of the bootstrap procedure, we also simulate the distribution of $\sup_{\bm{x}}|\vQn(\bm{x},\widehat{\bm{\theta}})|$ via Monte Carlo; that is, the data are generated from $Q_{\bm{\theta}}(\bm{x})$ (instead of $Q_{\widehat{\bm{\theta}}_{obs}}(\bm{x})$ as in the parametric bootstrap) and  the estimation process is repeated at each replicate. In all the three cases, the supremum is taken over a grid of $2000$ equidistant points  over $\mathcal{X}$.
 The results obtained are shown in Figure \ref{appQ}. 
The three simulated distributions are effectively overlapping, providing evidence that the parametric boostrap does recover the distribution of $\sup_{\bm{x}}|\vQn(\bm{x},\widehat{\bm{\theta}})|$. Not surprisingly, this is true even when relying on $\widetilde{v}_{Q,n}(\bm{x},{\bm{\theta}})$ instead of $\vQn(\bm{x},\widehat{\bm{\theta}})$ due to the small error associated with approximating the latter with the former. Notice that, this is true even if our sample size is limited to 100 observations.  Moreover,  working with the projected empirical process, $\widetilde{v}_{Q,n}(\bm{x},{\bm{\theta}})$, provides a remarkable computational gain compared to $\vQn(\bm{x},\widehat{\bm{\theta}})$. As shown in Table \ref{appQtab}, simulating the distribution of $\sup_{\bm{x}}|\vQn(\bm{x},\widehat{\bm{\theta}})|$ using $10,000$ replicates required approximately $12$ hours of (user+system) CPU time, whereas simulating the distribution of $\sup_{\bm{x}}|\widetilde{v}_{Q,n}(\bm{x},{\bm{\theta}})|$ required $9.5$ minutes. 
\section{Connecting tests for $F$ and tests for $Q$}
\label{FtoQ}
 In principle, we could proceed testing any $F_{\bm{\beta}}\not\equiv Q_{\bm{\theta}}$ following exactly the same steps described in Section \ref{testingQ}. In many practical situations, however,  $F_{\bm{\beta}}$ may be sufficiently complex to make the evaluation of the score functions over several samples impractical. To overcome this limitation, we proceed by  constructing  a new set of test statistics, namely $\widetilde{D}_{F}$, $\widetilde{\omega}^2_F$, and $\widetilde{A}^2_F$, whose limiting distributions, under $F_{\bm{\beta}}$, are the same as those of $\widehat{D}_Q$, $\widehat{\omega}^2_Q$, and $\widehat{A}^2_Q$ in \eqref{DQ}, under $Q_{\bm{\theta}}$. As a result, one can compute $\widetilde{D}_{F}$, $\widetilde{\omega}^2_F$, and $\widetilde{A}^2_F$ only once on  the data observed, and compare their values with the simulated distribution of $\widehat{D} _{Q}$, $\widehat{\omega}^2_Q$, and $\widehat{A}^2_Q$. This can be done by means of the K-2  transform  \cite{khm16,khm17} as described below.

Let $\bm{\beta}\in \Real^p$ be the vector of unknown parameters characterizing $F_{\bm{\beta}}$, let $f_{\bm{\beta}}(\bm{x})$ be its density (either pdf or pmf) and denote with $a_{\bm{\beta}_j}$, $j=1,\dots,p$, its normalized score functions. The latter can be constructed as in \eqref{scoreb} by replacing $q_{\bm{\theta}}$ and $\bm{\theta}$ with  $f_{\bm{\beta}}$ and $\bm{\beta}$, respectively.
For what follows, we require that  $f_{\bm{\beta}}( \bm{x})=0$ if and only if  $q_{\bm{\theta}}( \bm{x} )=0$, that is, the two densities must share the same support. Moreover, we assume that $\bm{\beta}$ and $\bm{\theta}$, have the same dimension $p$.
 
Equations \eqref{psihat}-\eqref{projection1a} imply that the process $\widetilde{v}_{Q,n}(\bm{x},{\bm{\theta}})$ ``lives'' in the space of functions  ${\mathcal{L}}_{\perp}(Q_{\bm{\theta}})$ such that
\begin{align}
\label{LQper}
{\mathcal{L}}_{\perp}(Q_{\bm{\theta}})=\bigl\{&\widetilde{\psi}: {\langle \widetilde{\psi},\widetilde{\psi}\rangle} _{Q_{\bm{\theta}}}<\infty\\
&{\langle \widetilde{\psi},1\rangle} _{Q_{\bm{\theta}}}=0,\quad \text{and}\\
&{\langle\widetilde{\psi},b_{\bm{\theta}_j}\rangle} _{Q_{\bm{\theta}}}=0, \quad\text{for all }j=1,\dots,p\bigl\}
\end{align}
That is, each function in ${\mathcal{L}}_{\perp}(Q_{\bm{\theta}})$ is square-integrable with respect to $Q_{\bm{\theta}}$, has mean zero, and   is orthogonal to the normalized score functions  $b_{\bm{\theta}_j}$, $j=1,\dots,p$, under $Q_{\bm{\theta}}$.
Moreover, one can  show that, under $Q_{\bm{\theta}}$, the process $\widetilde{v}_n(\bm{x},{\bm{\theta}})$ is asymptotically Gaussian  with mean $\langle\widetilde{\psi}_{\bm{x},\bm{\theta}},1\rangle_{Q_{\bm{\theta}}}=0$ and covariance $\langle\widetilde{\psi}_{\bm{x},\bm{\theta}},\widetilde{\psi}_{\bm{s},\bm{\theta}}\rangle_{Q_{\bm{\theta}}}<\infty$. 

The rationale behind the K-2  transformation is that of constructing a suitable map  which allows us to transform functions  $\widetilde{\psi}_{\bm{x},\bm{\theta}}\in\mathcal{L}_{\perp}({Q_{\bm{\theta}}})$ into functions  in ${\mathcal{L}}_{\perp}(F_{\bm{\beta}})$, i.e.,
\begin{align}
\label{LFper}
{\mathcal{L}}_{\perp}(F_{\bm{\beta}})=\bigl\{&\widetilde{\phi}: {\langle \widetilde{\phi},\widetilde{\phi}\rangle}_{F_{\bm{\beta}}}<\infty,\\
&{\langle \widetilde{\phi},1\rangle}_{F_{\bm{\beta}}}=0,\quad\text{and}\\
&{\langle\widetilde{\phi},a_j\rangle}_{F_{\bm{\beta}}}=0, \quad\text{for all }j=1,\dots,p\bigl\},
\end{align}
where $\langle\cdot,\cdot\rangle_{F_{\bm{\beta}}}$ can be defined similarly to $\langle\cdot,\cdot\rangle_{Q_{\bm{\theta}}}$ in \eqref{inner1}. Notice that  ${\mathcal{L}}_{\perp}(F_{\bm{\beta}})\subset \mathcal{L}(F_{\bm{\beta}}) \subset L^2(F_{\bm{\beta}})$, with
\begin{align*}
\quad L^2(F_{\bm{\beta}})&=\bigl\{\widetilde{\phi}: {\langle \widetilde{\phi},\widetilde{\phi}\rangle}_{F_{\bm{\beta}}}<\infty\bigl\},\quad\text{and}\\
\mathcal{L}(F_{\bm{\beta}})&=\bigl\{\widetilde{\phi}: {\langle\widetilde{\phi},1\rangle}_{F_{\bm{\beta}}}=0,\quad {\langle \widetilde{\phi},\widetilde{\phi}\rangle}_{F_{\bm{\beta}}}<\infty\bigl\}.
\end{align*}

 It follows that, for suitable choices of $\widetilde{\phi}$, namely $\widetilde{\phi}_{\bm{x},\bm{\lambda}}$  (soon to be defined),  the   process $\widetilde{v}_{Q,n}(\bm{x},{\bm{\theta}})$ in \eqref{projection1a} and the empirical process
\begin{align}
\label{vnF}
\widetilde{v}_{F,n}(\bm{x},\bm{\lambda})=\frac{1}{\sqrt{n}}\sum_{i=1}^n\widetilde{\phi}_{\bm{x},\bm{\lambda}}(\bm{x}_i),\quad  \text{with $\bm{\lambda}=(\bm{\theta},\bm{\beta})$,}
\end{align}
 have the same asymptotic distribution (under $Q_{\bm{\theta}}$ and $F_{\bm{\beta}}$, respectively). 
 Specifically, in virtue of Gaussianity, we can fully characterize the distribution of  $\widetilde{v}_{F,n}(\bm{x},{\bm{\theta}})$ and $\widetilde{v}_{Q,n}(\bm{x},\bm{\lambda})$ considering only their mean and covariance.  Therefore, to achieve our purpose, it is sufficient  to identify a set of functions $\widetilde{\phi}_{\bm{x},\bm{\lambda}}\in\mathcal{L}_{\perp}(F_{\bm{\beta}})$ such that the mean and covariance functions of  $\widetilde{v}_{F,n}(\bm{x},{\bm{\theta}})$ and $\widetilde{v}_{Q,n}(\bm{x},\bm{\lambda})$ are the same, i.e.,
\begin{align*}
\langle\widetilde{\psi}_{\bm{x},\bm{\theta}},1\rangle_{Q_{\bm{\theta}}}&=\langle\widetilde{\phi}_{\bm{x},\bm{\lambda}},1\rangle_{F_{\bm{\beta}}}=0\quad\text{and}\\
\langle\widetilde{\psi}_{\bm{x},\bm{\theta}},\widetilde{\psi}_{\bm{s},\bm{\theta}}\rangle_{Q_{\bm{\theta}}}&=\langle\widetilde{\phi}_{\bm{x},\bm{\lambda}},\widetilde{\phi}_{\bm{s},\bm{\lambda}}\rangle_{F_{\bm{\beta}}}
\end{align*}
The functions $\widetilde{\phi}_{\bm{x},\bm{\lambda}}$  can be constructed as outlined below.


\textbf{Step 1 -}  Map  the functions $\psi_{\bm{x},\bm{\theta}}$ in equation \eqref{phi} and the normalized score functions $b_{\bm{\theta}_j}$ into $L^2(F_{\bm{\beta}})$ via the isometry 
\[l_{}(\bm{x})=\sqrt{\frac{q_{\bm{\theta}}(\bm{x})}{f_{\bm{\beta}}(\bm{x})}}.\] 
Obtain  
\begin{align}
\label{f0}
l_{}(\bm{t})\psi_{\bm{x},\bm{\theta}}(\bm{t})&\in  L^2(F_{\bm{\beta}})\quad\text{and}\\
\label{f1}
l_{}(\bm{t})b_{\bm{\theta}_j}(\bm{t})&\in  L^2(F_{\bm{\beta}}). 
 \end{align}
 For instance, to see \eqref{f0}, consider the inner product
  \begin{align*}
 \langle l_{}{\psi}_{\bm{x},\bm{\theta}},& l_{}{\psi}_{\bm{x},\bm{\theta}}\rangle_{F_{\bm{\beta}}}=\int_{\mathcal{X}} l_{}^2(\bm{t}){\psi}^2_{\bm{x},\bm{\theta}}(\bm{t})f_{\bm{\beta}}(\bm{t})\text{d}\bm{t}\\
 &=\int_{\mathcal{X}} \frac{q_{\bm{\theta}}(\bm{t})}{f_{\bm{\beta}}(\bm{t})}{\psi}^2_{\bm{x},\bm{\theta}}(\bm{t})f_{\bm{\beta}}(\bm{t})\text{d}\bm{t}\\
 &=\int_{\mathcal{X}} {\psi}^2_{\bm{x},\bm{\theta}}(\bm{t})q_{\bm{\theta}}(\bm{t})\text{d}\bm{t}= \langle {\psi}_{\bm{x},\bm{\theta}}, {\psi}_{\bm{x},\bm{\theta}}\rangle_Q<\infty. 
 \end{align*}
Equivalent calculations can be used to show \eqref{f1}.

\textbf{Step 2 -} Map the functions in \eqref{f0} and \eqref{f1} into $\mathcal{L}(F_{\bm{\beta}})$ by means of the unitary operator\footnote{A unitary operator is an operator that preserves the inner product. That is, if an operator $K$ is unitary in the Hilbert space $\mathcal{H}$ equipped with the inner product $\langle,\rangle_{\mathcal{H}}$, then $\langle Kh_1,Kh_2\rangle_{\mathcal{H}}=\langle h_1,h_2\rangle_{\mathcal{H}}$, for every $h_1,h_2\in\mathcal{H}$.}, $K$, and defined as
\begin{equation}
\label{K}
Kh(\bm{t})=h(\bm{t})-\frac{1-l_{}(\bm{t})}{1-\langle l_{},1\rangle_{F_{\bm{\beta}}}}\langle  1-l_{},h \rangle_{F_{\bm{\beta}}}, 
\end{equation}
where the notation $Kh(\bm{t})$ is used to indicate that the operator $K$ acts on everything on its right. Obtain
\begin{align}
\label{k0}
Kl_{}(\bm{t})\psi_{\bm{x},\bm{\theta}}(\bm{t})&\in \mathcal{L}(F_{\bm{\beta}})\\
\label{k1}
\text{and}\quad c_{\bm{\lambda}_j}(\bm{t})=Kl_{}(\bm{t})b_{\bm{\theta}_j}(\bm{t})&\in \mathcal{L}(F_{\bm{\beta}}).
\end{align}
To see  \eqref{k0}, write
{\fontsize{9.5}{9.5}\selectfont{
\begin{align*}
Kl_{}(\bm{t})\psi_{\bm{x},\bm{\theta}}(\bm{t})&=l_{}(\bm{t})\psi_{\bm{x},\bm{\theta}}(\bm{t})-\\
&\frac{1-l_{}(\bm{x})}{1-\int_{\mathcal{X}} l_{}(\bm{t})f_{\bm{\beta}}(\bm{t})\text{d}\bm{t}}\int_{\mathcal{X}} l_{}(t)\psi_{\bm{x},\bm{\theta}}(\bm{t})f_{\bm{\beta}}(\bm{t}) \text{d}\bm{t}.
\end{align*}}}
It follows that
{\fontsize{9.5}{9.5}\selectfont{
\begin{align*}
 \langle K&l_{}{\psi}_{\bm{x},\bm{\theta}},1\rangle_F=\int_{\mathcal{X}} l_{}(\bm{x})\psi_{\bm{x},\bm{\theta}}(\bm{t})f_{\bm{\beta}}(\bm{t}) \text{d}\bm{t}-\\
 &\frac{1-\int_{\mathcal{X}} l_{}(\bm{t})f_{\bm{\beta}}(\bm{t})\text{d}\bm{t}}{1-\int_{\mathcal{X}} l_{}(\bm{t})f_{\bm{\beta}}(\bm{t})\text{d}\bm{t}}\int_{\mathcal{X}} l_{}(\bm{t})\psi_{\bm{x},\bm{\theta}}(\bm{t}) f_{\bm{\beta}}(\bm{t})\text{d}\bm{t}=0.
\end{align*}}}
One can proceed similarly for \eqref{k1}.

\textbf{Step 3 -} Map each function $c_{\bm{\lambda}_j}$ in \eqref{k1} with $j>1$ into functions $\tilde{c}_{\bm{\lambda}_j}$ orthogonal to each $a_{\bm{\beta}_k}$ with $k<j$. This can be done by means of  the unitary operator 
\begin{equation}
\label{Uj}
U_{a_{\bm{\beta}_j}c_{\bm{\lambda}_j}}h(\bm{t})=h(\bm{t})-\frac{\langle a_{\bm{\beta}_j}-c_{\bm{\lambda}_j},\cdot \rangle_{F_{\bm{\beta}}}}{1-\langle a_{\bm{\beta}_j},c_{\bm{\lambda}_j}\rangle_{F_{\bm{\beta}}}}(a_{\bm{\beta}_j}(\bm{t})-c_{\bm{\lambda}_j}(\bm{t})).
\end{equation}
One can easily verify that the operator $U_{a_{\bm{\beta}_j}c_{\bm{\lambda}_j}}$ maps the functions $a_{\bm{\beta}_j}$ into functions $c_{\bm{\lambda}_j}$, and vice-versa, whereas, it leaves functions orthogonal to both $a_{\bm{\beta}_j}$ and $c_{\bm{\lambda}_j}$ unchanged. 

We construct $\tilde{c}_2,\dots,\tilde{c}_p$, by combining operators of the form in \eqref{Uj}, i.e.,
\begin{equation}
\begin{split}
\label{ctildes}
\tilde{c}_{\bm{\lambda}_2}(\bm{t})&=U_{a_{\bm{\beta}_1}c_{\bm{\lambda}_1}}c_{\bm{\lambda}_2}(\bm{t})\\
\tilde{c}_{\bm{\lambda}_3}(\bm{t})&=U_{a_{\bm{\beta}_2}\tilde{c}_{\bm{\lambda}_2}}U_{a_{\bm{\beta}_1}c_{\bm{\lambda}_1}}c_{\bm{\lambda}_3}(\bm{t})\\
&\dots\\
\tilde{c}_{\bm{\lambda}_p}(\bm{t})&=U_{a_{\bm{\beta}_{(p-1)}}\tilde{c}_{\bm{\lambda}_{(p-1)}}}\dots U_{a_{\bm{\beta}_1}c_{\bm{\lambda}_1}}c_{\bm{\lambda}_p}(\bm{t}),
\end{split}
\end{equation}
where each operator $U_{a_{\bm{\beta}_j}c_{\bm{\lambda}_j}}$ acts on everything on its right. As highlighted in what follows, these functions are needed to rotate $c_{\bm{\lambda}_j}$s into $a_{\bm{\beta}_{j}}$s.

\textbf{Step 4 -} Consider the unitary operator
\begin{equation}
\label{U}
\bm{U}h(\bm{t})=U_{a_{\bm{\beta}_p}\tilde{c}_{\bm{\lambda}_p}}\dots U_{a_{\bm{\beta}_2}\tilde{c}_{\bm{\lambda}_2}}U_{a_{\bm{\beta}_1}c_{\bm{\lambda}_1}}h(\bm{t})
\end{equation}
and set
\begin{equation}
\label{phis}
\phi_{\bm{x},\bm{\lambda}}(\bm{t})=\bm{U}Kl_{}(\bm{t}){\psi}_{\bm{x},\bm{\theta}}(\bm{t})
\end{equation}
Map   each $c_{\bm{\lambda}_j}$ into $a_{\bm{\beta}_j}$ via $\bm{U}$
and  apply the latter to  $Kl_{}\widetilde{\psi}_{\bm{x},\bm{\theta}}$. Obtain $\widetilde{\phi}_{\bm{x},\bm{\lambda}}\in\mathcal{L}_{\perp}(F_{\bm{\beta}})$ such that
\begin{align}
\label{finalphi}
&\widetilde{\phi}_{\bm{x},\bm{\lambda}}(\bm{t})=\bm{U}K l_{}(\bm{t})\widetilde{\psi}_{\bm{x},\bm{\theta}}(\bm{t})\\
\label{finalphi1}
& =\bm{U}K\Bigl[l_{}(\bm{t})\psi_{\bm{x},\bm{\theta}}(\bm{t})-\sum_{j=1}^p l_{}(\bm{t}) b_{\bm{\theta}_j}(\bm{t}) {\langle l_{}b_{\bm{\theta}_j}, l_{}\psi_{\bm{x},\bm{\theta}}\rangle}_{F_{\bm{\beta}}}\Bigl]\\
\label{finalphi2}
& =\bm{U}\Bigl[Kl_{}(\bm{t})\psi_{\bm{x},\bm{\theta}}(\bm{t})-\sum_{j=1}^p c_{\bm{\lambda j}} {\langle c_{\bm{\lambda}_j}, Kl_{}\psi_{\bm{x},\bm{\theta}}\rangle}_{F_{\bm{\beta}}}\Bigl]\\
\label{finalphi3}
&=\phi_{\bm{x},\bm{\lambda}}(\bm{t})-\sum_{j=1}a_{\bm{\beta}_j}(\bm{t}){\langle a_{\bm{\beta}_j}, \phi_{\bm{x},\bm{\lambda}}\rangle}_{F_{\bm{\beta}}}.
\end{align}
Where \eqref{finalphi1} follows from the definition of the functions $\widetilde{\psi}_{\bm{x},\bm{\theta}}$ in \eqref{psihat}. Equation \eqref{finalphi2} follow from \eqref{k1}, from the fact that  $K$ is unitary (and thus it preserve the inner product), and because the isometry $l_{}$ is such that  $\langle l_{}h,  l_{}h\rangle_{F_{\bm{\beta}}}=\langle h,  h\rangle_{Q_{\bm{\theta}}}$. Equation \eqref{finalphi3} follows from \eqref{phis} and the properties of the operator $\bm{U}$ (that is, it is unitary and it maps each $c_{\bm{\lambda}_j}$ into $a_{\bm{\beta}_j}$). 
To see the latter, consider for instance $\bm{U}c_{\bm{\lambda}_1}(\bm{t})$, i.e.,
\begin{align}
\label{e0}
\bm{U}c_{\bm{\lambda}_1}(\bm{t})=&U_{a_{\bm{\beta}_p}\tilde{c}_{\bm{\lambda}_p}}\dots U_{a_{\bm{\beta}_2}\tilde{c}_{\bm{\lambda}_2}}U_{a_{\bm{\beta}_1}c_{\bm{\lambda}_1}}c_{\bm{\lambda}_1}(\bm{t})\\
\label{e1}
=&U_{a_{\bm{\beta}_p}\tilde{c}_{\bm{\lambda}_p}}\dots U_{a_{\bm{\beta}_2}\tilde{c}_{\bm{\lambda}_2}}a_{\bm{\beta}_1}(\bm{t})\\
\label{e2}
=&a_{\bm{\beta}_1}(\bm{t})
\end{align}
where \eqref{e1} follows since $U_{a_{\bm{\beta}_1}c_{\bm{\lambda}_1}}$ maps $c_{\bm{\lambda}_1}$ into $a_{\bm{\beta}_1}$. Whereas,  \eqref{e2} follows from the fact that each $a_{\bm{\beta}_j}$ and $\tilde{c}_{\bm{\lambda}_j}$, with $j\geq 2$, are orthogonal to $a_{\bm{\beta}_1}$ and each $U_{a_{\bm{\beta}_j}\tilde{c}_{\bm{\lambda}_j}}$ leaves functions orthogonal to $a_{\bm{\beta}_j}$ and $\tilde{c}_{\bm{\lambda}_j}$ unchanged.  
Moreover, to see that $\widetilde{\phi}_{\bm{x},\bm{\lambda}}=\bm{U}Kl_{}\widetilde{\psi}_{\bm{x},\bm{\theta}}\in \mathcal{L}_\perp (F_{\bm{\beta}})$, consider 
 \begin{align}
 \label{d1}
 \langle \bm{U}Kl_{}\widetilde{\psi}_{\bm{x},\bm{\theta}},a_{\bm{\beta}_j}\rangle_{F_{\bm{\beta}}}&= \langle \bm{U}Kl_{}\widetilde{\psi}_{\bm{x},\bm{\theta}},\bm{U}{c}_{\bm{\lambda}_j}\rangle_{F_{\bm{\beta}}}\\
  \label{d2}
 &=\langle Kl\widetilde{\psi}_{\bm{x},\bm{\theta}},{c}_{\bm{\lambda}_j}\rangle_{F_{\bm{\beta}}}\\
   \label{d3}
 &=\langle l_{}\widetilde{\psi}_{\bm{x},\bm{\theta}},lb_{\bm{\theta}_j}\rangle_{F_{\bm{\beta}}}\\
    &=\langle \widetilde{\psi}_{\bm{x},\bm{\theta}},b_{\bm{\theta}_j}\rangle_{Q_{\bm{\theta}}}=0,
 \end{align}
 where the equalities in \eqref{d2}-\eqref{d3} follow from the properties $\bm{U}$, $K$ and $l_{}$.

Clearly, for $Q_{\bm{\theta}}$ and $F_{\bm{\beta}}$ discrete, all the integrals involved in Steps 1-4 need to be replaced by summations over all the points of the search region $\mathcal{\bm{X}}$. Moreover, it should be noted that, in virtue of the properties of the $\bm{U}$, $K$ and $l_{}$ we have
\begin{align}
\langle b_{\bm{\theta}_j}, \widetilde{\psi}_{\bm{x},\bm{\theta}}\rangle_{Q_{\bm{\theta}}}= {\langle c_{\bm{\lambda}_j}, Kl_{}\psi_{\bm{x},\bm{\theta}}\rangle}_{F_{\bm{\beta}}}={\langle a_{\bm{\beta}_j}, \phi_{\bm{x},\bm{\lambda}}\rangle}_{F_{\bm{\beta}}}.
\end{align}
Hence, when evaluating the functions $\widetilde{\phi}_{\bm{x},\bm{\lambda}}(\bm{t})$ in \eqref{finalphi}, one can avoid computing ${\langle a_{\bm{\beta}_j}, \phi_{\bm{x},\bm{\lambda}}\rangle}_{F_{\bm{\beta}}}$ by replacing it with $\langle b_{\bm{\theta}_j},\widetilde{\psi}_{\bm{x},\bm{\theta}}\rangle_{Q_{\bm{\theta}}}$.

From \eqref{finalphi}, it is easy to see that K-2  effectively consists of a combination of the unitary opertors $\bm{U}$, $K$ and the isometry $l_{}$.
Intuitively, in Step 1, the isometry $l_{}$ allows us to convert our functions ${\psi}_{\bm{x},\bm{\theta}}$, square-integrable in $Q_{\bm{\theta}}$, into square integrable functions in $F_{\bm{\beta}}$. The resulting functions $l_{}\psi_{\bm{x},\bm{\theta}}$ and $l_{\bm{\lambda}}b_{\theta j}$, however, do not have zero-mean with respect to $F_{\bm{\beta}}$  (they are not orthogonal to one). Therefore, in Step 2,  we apply the unitary operator $K$.  This brings us to the space $\mathcal{L}(F_{\bm{\beta}})$. If $\bm{\theta}$ and $\bm{\beta}$ were known, that is, if the two models were fully specified, the isometry $l_{}$ and the operator $K$ would only need to be applied to the functions $\psi_{\bm{x},\bm{\theta}}$ (as there would be no score functions) and no further mapping would be needed. Whereas, for $\bm{\theta}$ and $\bm{\beta}$ unknown, two extra steps  are neessary. That is because, in this setting, $\mathcal{L}(F_{\bm{\beta}})$ is not quite yet be in the space we want to be (i.e., $\mathcal{L}_{\perp}(F_{\bm{\beta}})$) as we have not yet achieved orthogonality with respect to the score functions $a_{\bm{\beta}_j}$s.  Hence, in Step 3, we exploit the unitary operator $\bm{U}$ to map our $c_{\bm{\lambda}_j}=Klb_{\theta j}$ into $\tilde{c}_{\bm{\lambda}_j}$ functions which are orthogonal to the $a_{\bm{\beta}_j}$. Finally, in Step 4, we rotate  the ${c}_{\bm{\lambda}j}$s into $a_{\bm{\beta}_j}$s via $\bm{U}$. The same operator is applied also to the functions $Kl_{}\psi_{\bm{x},\bm{\theta}}$ to ensure that the functions $\widetilde{\phi}_{\bm{x}\bm{\lambda}}=Kl\widetilde{\psi}_{\bm{x},\bm{\theta}}$ in \eqref{finalphi} are in $\mathcal{L}_\perp(F_{\bm{\beta}})$.

To test the hypothesis $H_0:P=F_{\bm{\beta}}$, we consider the K-2  rotated equivalent of the test statistics in \eqref{DQ}, i.e.,
\begin{equation}
\begin{split}
\label{DF}
\widetilde{D}_F&=\sup_{\bm{x}}|\widetilde{v}_{F,n}(\bm{x},\bm{\lambda})|,\quad \widetilde{\omega}^2_F=\int_{\mathcal{X}}\widetilde{v}^2_{F,n}(\bm{x},\bm{\lambda}) q_{\bm{\theta}}(\bm{x})\text{d}\bm{x}, \\
\text{and}&\quad \widetilde{A}^2_F=\int_{\mathcal{X}}\widetilde{v}^2_{F,n}(\bm{x},\bm{\lambda}) w_{\bm{\theta}}(\bm{x})q_{\bm{\theta}}(\bm{x})\text{d}\bm{x}
\end{split}
\end{equation}
with $\widetilde{v}_{F,n}(\bm{x},\bm{\lambda})$ as in \eqref{vnF}. Under $F_{\bm{\beta}}$ and $Q_{\bm{\theta}}$, respectively, $\widetilde{v}_{F,n}(\bm{x},\bm{\lambda})$ and $\widetilde{v}_{Q,n}(\bm{x},{\bm{\theta}})$ have the same asymptotic distribution, and  the same is true for the statistics in \eqref{DQ} and \eqref{DF}.


 \begin{figure*}[htb]
\begin{tabular*}{\textwidth}{@{\extracolsep{\fill}}@{}c@{}c@{}c@{}}
      \includegraphics[width=60mm]{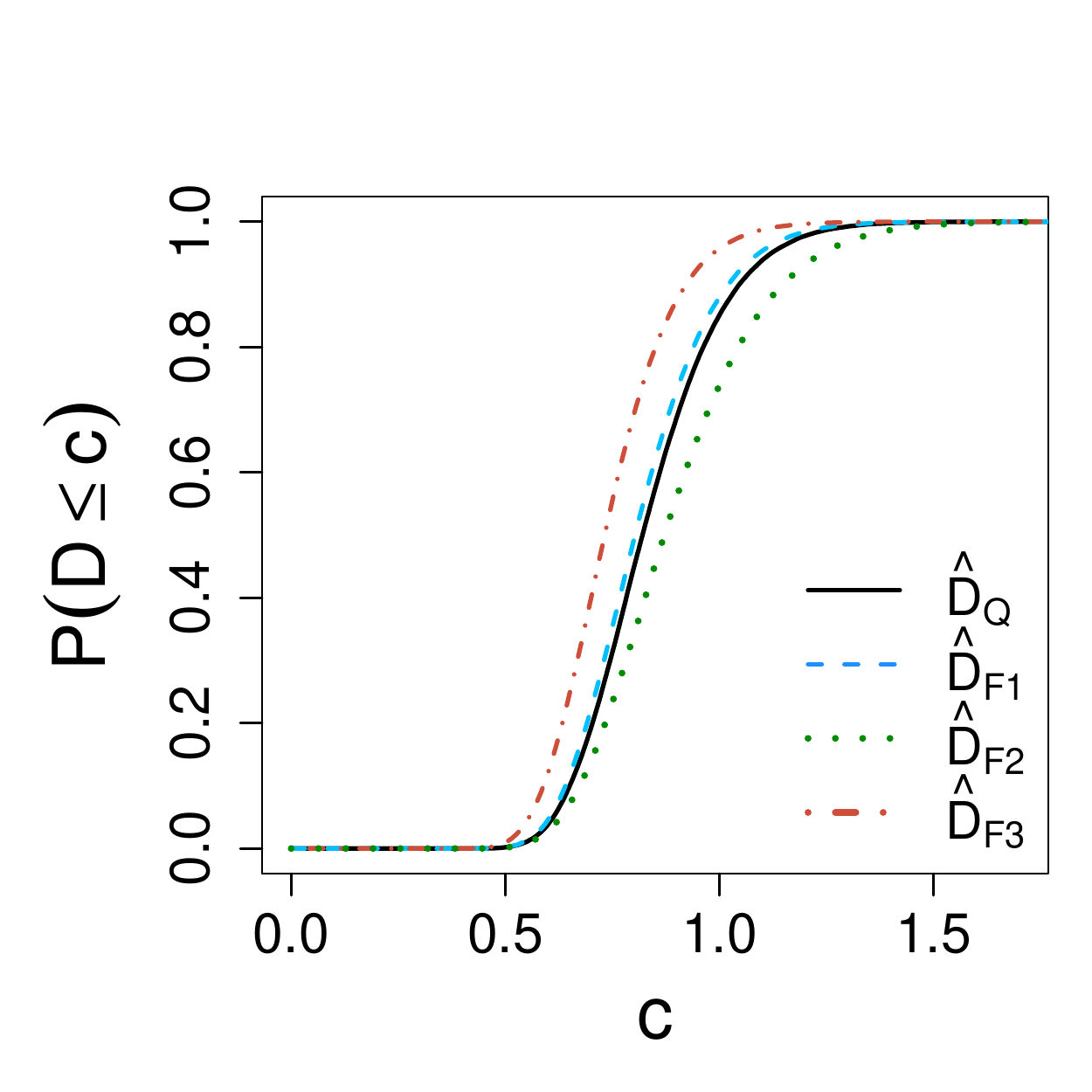} &\hspace{-2cm}   \includegraphics[width=60mm]{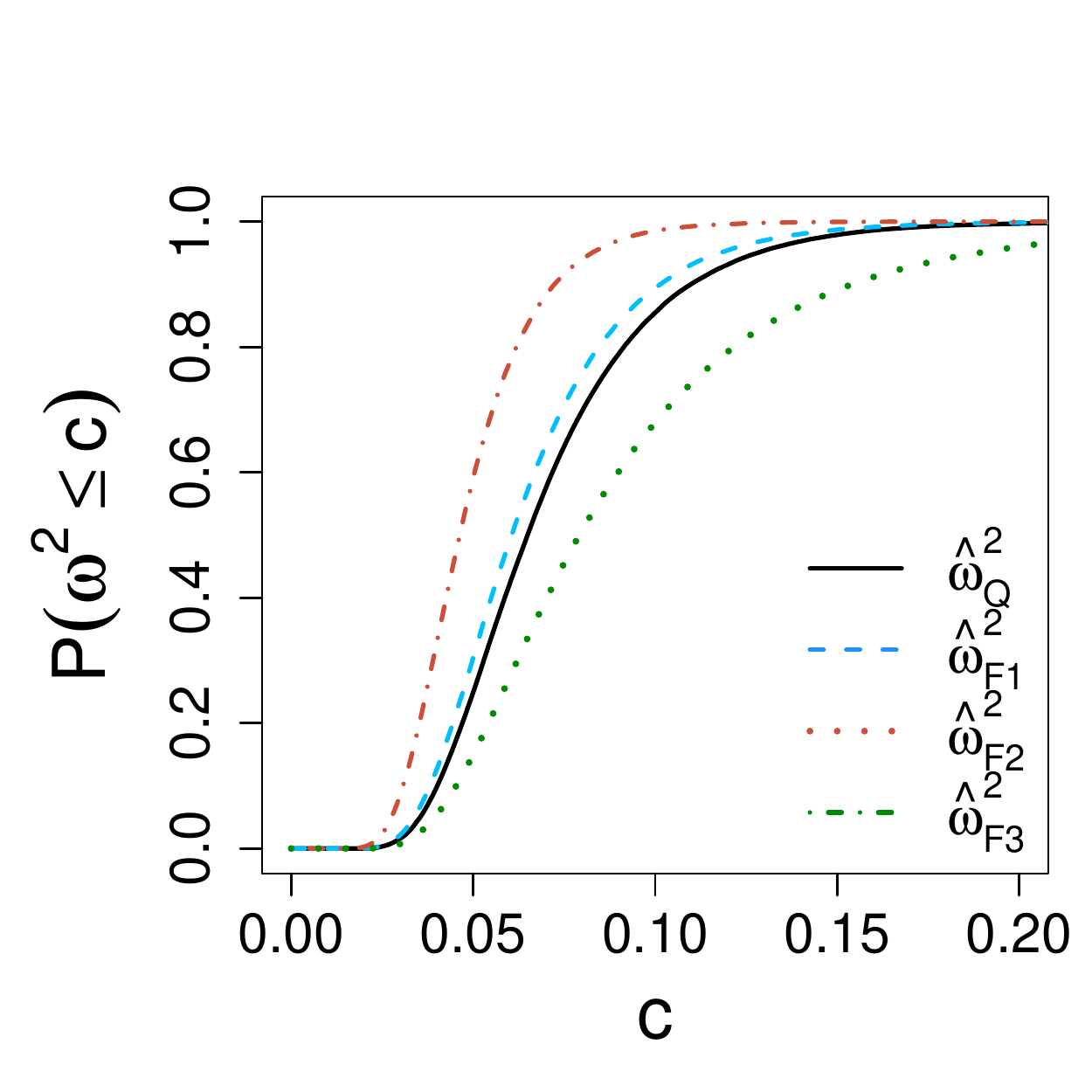}&  \hspace{-2cm} \includegraphics[width=60mm]{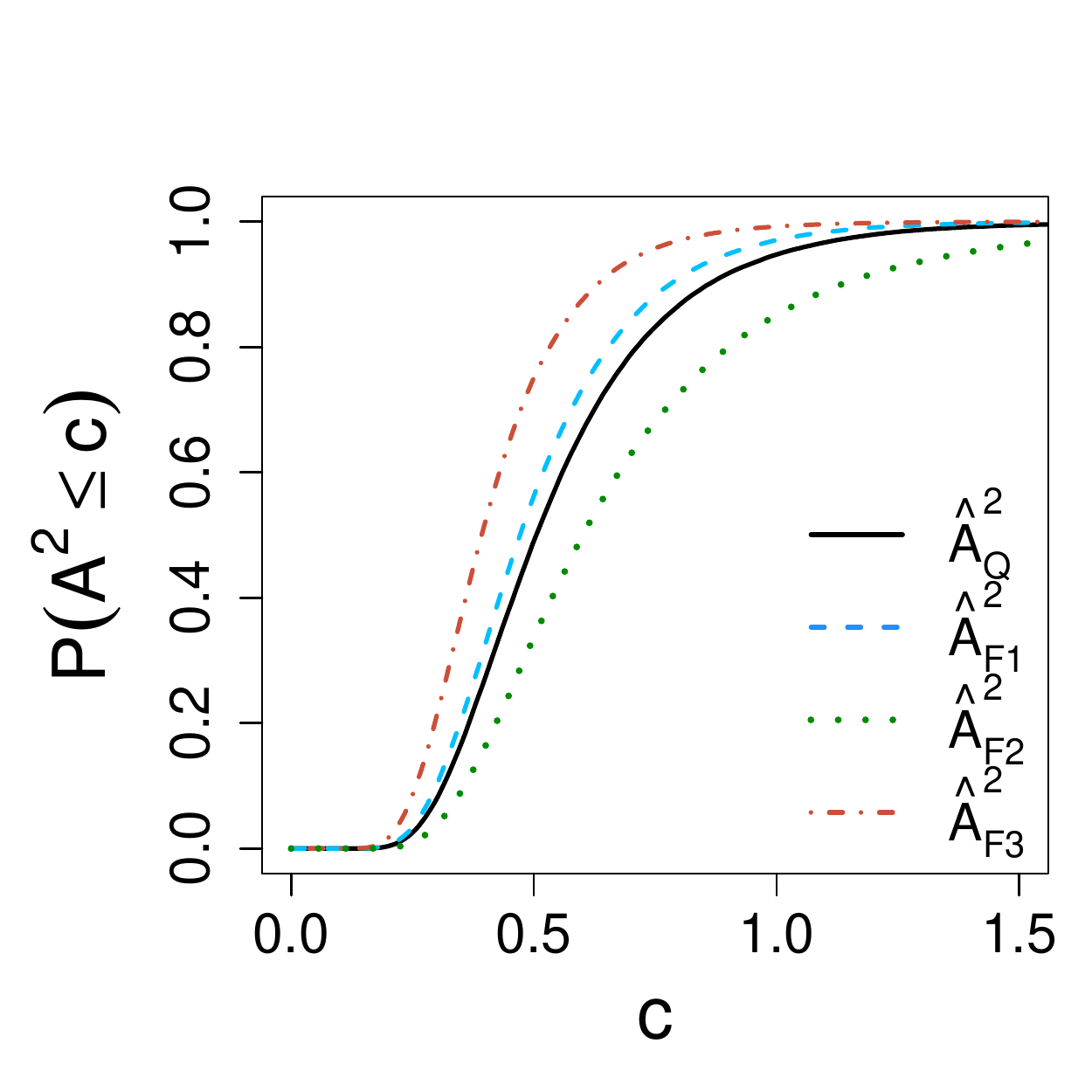}\\[-1.2cm]
      \includegraphics[width=60mm]{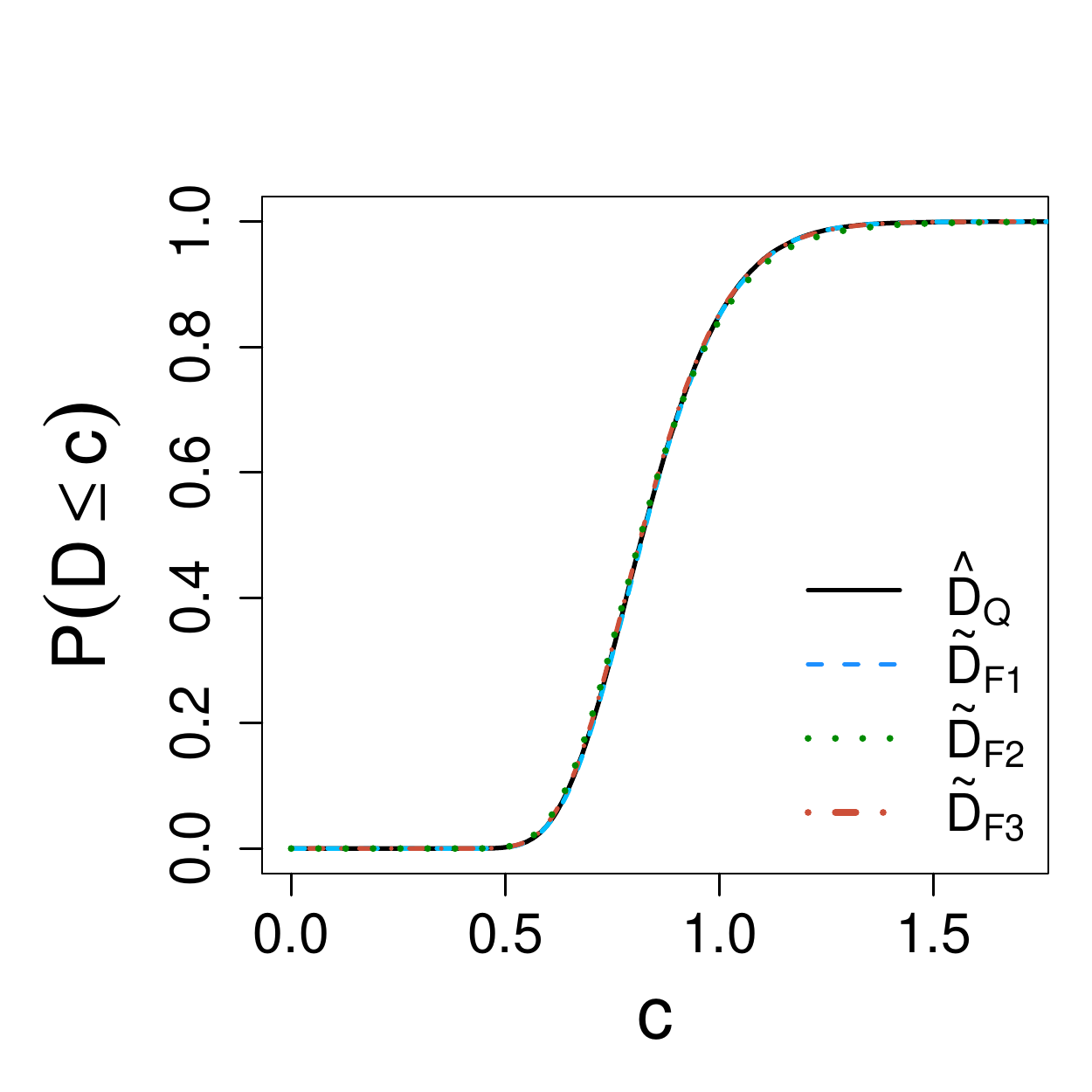} &\hspace{-2cm}   \includegraphics[width=60mm]{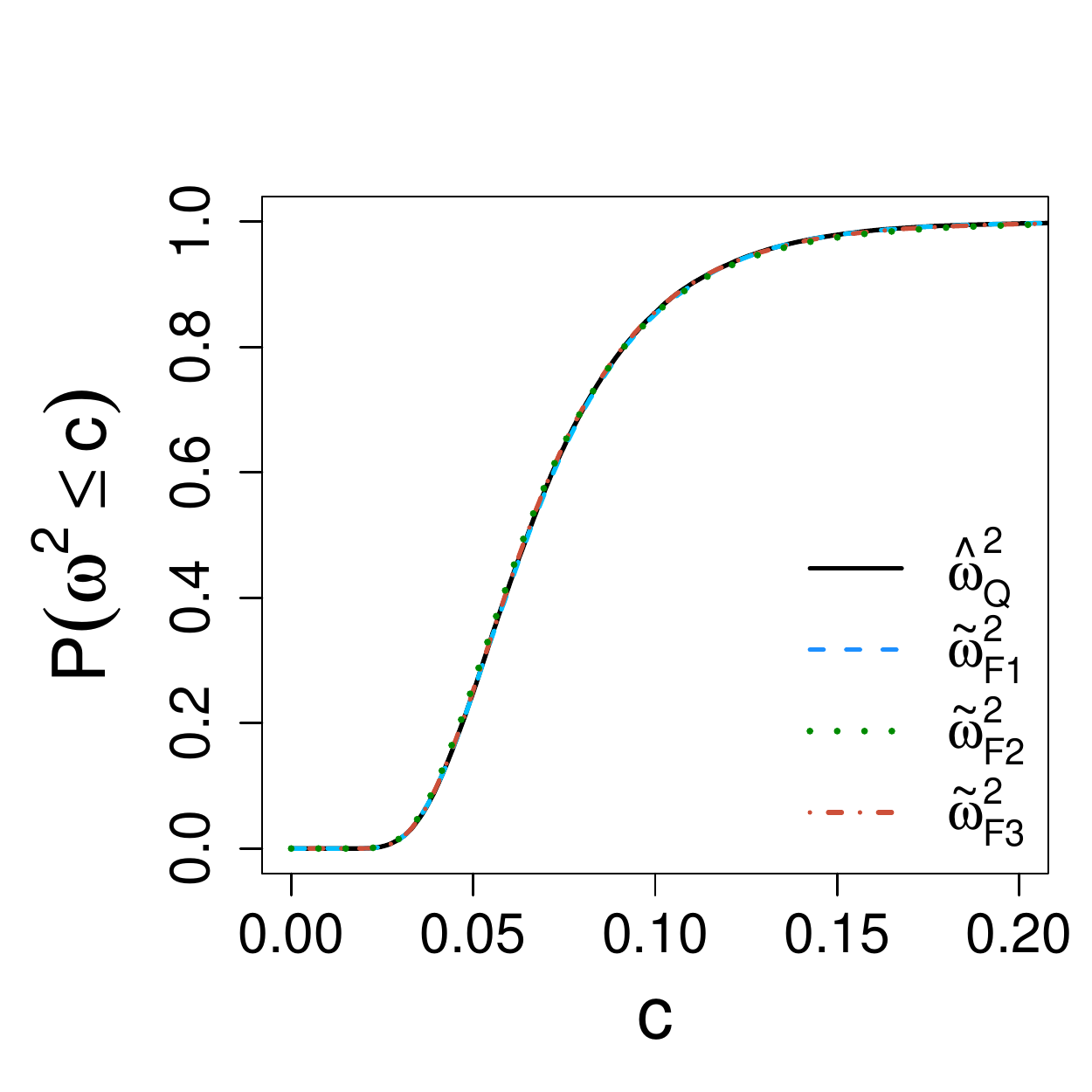}&  \hspace{-2cm} \includegraphics[width=60mm]{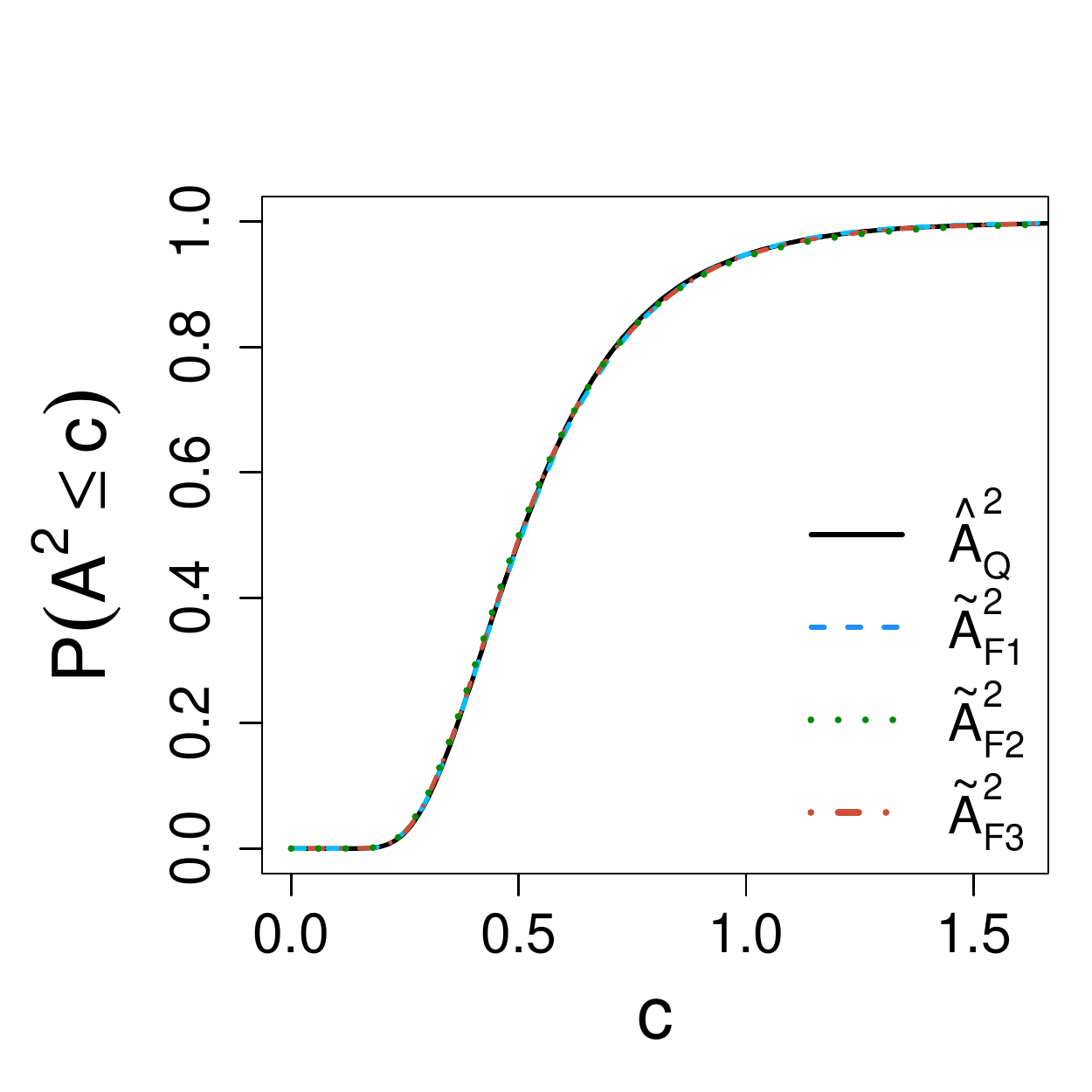}\\[-0.5cm]
\end{tabular*}
\caption[Figure 2]{Upper panels: Comparing the simulated null distributions of the test statistics in \eqref{DQ} for $q$ in \eqref{q} and for each candidate model $f_m$, $m=1,\dots,3$, in \eqref{fs}.  Bottom panels: Comparing the simulated null distributions of the test statistics in \eqref{DQ} for $q$ with the K-2 rotated statistics in \eqref{DF}
 for each  $f_m$, $m=1,\dots,3$. Each simulation involves 100,000 replicates and $100$ observations.}
\label{Fig2}
\end{figure*}
 \begin{table*}[htb]
\fontsize{9}{9}\selectfont
\begin{tabular}{!{\vrule width 1.5pt}c!{\vrule width 1.4pt}cccccc!{\vrule width 1.5pt}cccccc!{\vrule width 1.5pt}cccccc!{\vrule width 1.5pt}}
                                 \noalign{\hrule height 1.5pt}

                  &&&&&&  &&&&&& &&&&&&\\[-1ex]
                                   &\multicolumn{6}{c!{\vrule width 1.5pt}}{$\alpha=0.001$} &\multicolumn{6}{c!{\vrule width 1.5pt}}{$\alpha=0.05$}&\multicolumn{6}{c!{\vrule width 1.5pt}}{$\alpha=0.1$}\\[-1ex]
                                                   &&&& &&  &&&& &&&&&& &&\\[-1ex]
\hline
    &&&   &  \multicolumn{1}{|c}{}&&&&&   &  \multicolumn{1}{|c}{}&&&&&   &  \multicolumn{1}{|c}{}&&\\[-1.5ex]
   Null Distribution &$\widehat{D}$ & $\widehat{\omega}^2$ & $\widehat{A}^2$ &\multicolumn{1}{|c}{$\widetilde{D}$}&$\widetilde{\omega}^2$ &$\widetilde{A}^2$&$\widehat{D}$ & $\widehat{\omega}^2$ & $\widehat{A}^2$ &\multicolumn{1}{|c}{$\widetilde{D}$}&$\widetilde{\omega}^2$ &$\widetilde{A}^2$&$\widehat{D}$ & $\widehat{\omega}^2$ & $\widehat{A}^2$ &\multicolumn{1}{|c}{$\widetilde{D}$}&$\widetilde{\omega}^2$ &$\widetilde{A}^2$\\[-0.8ex]
                       &&&&\multicolumn{3}{|c!{\vrule width 1.5pt}}{(K-2  rotated)}&&&&\multicolumn{3}{|c!{\vrule width 1.5pt}}{(K-2  rotated)}&&&&\multicolumn{3}{|c!{\vrule width 1.5pt}}{(K-2  rotated)}\\[-0.5ex]
                                 \noalign{\hrule height 1.5pt}
                       &&&   &  \multicolumn{1}{|c}{}&&&&&   &  \multicolumn{1}{|c}{}&&&&&   &  \multicolumn{1}{|c}{}&&\\[-1.5ex]
 $Q$    & .4773 & .7785& .4633  &  \multicolumn{1}{|c}{-} &-&-& .9331 & .9817& .9382  &  \multicolumn{1}{|c}{-} &-&-& .9679 & .9914& .9722  &  \multicolumn{1}{|c}{-} &-&- \\
 $F_1$  &.3872  &.6762&  .4815 &\multicolumn{1}{|c}{.1578}&1& 1& .8623&.9529&.9092  &\multicolumn{1}{|c}{.6971}&1& 1&.9221 &.9748&.9505    &\multicolumn{1}{|c}{.8086}&1& 1\\
$F_2$&.0036&.0025 & .0053 &\multicolumn{1}{|c}{.0058}&.0226&.0156&.1078&.1019&.1237&\multicolumn{1}{|c}{.1336}&.2422&.2541&.1876&.185&.2127&\multicolumn{1}{|c}{.2233}&.3618&.3770\\
 $F_3$&.6452&.7947&.0295 &\multicolumn{1}{|c}{.5062}&.7975&.6036&.9528&.9820&.6356&\multicolumn{1}{|c}{.9153}&.9746&.9470&.9757&.9915&.7974&\multicolumn{1}{|c}{.9543}&.9874&.9730\\   [-1.5ex]
                       &&& & \multicolumn{1}{|c}{}   &&&&& & \multicolumn{1}{|c}{}   &&&&& & \multicolumn{1}{|c}{}   &&\\
  
                                                    \noalign{\hrule height 1.4pt}
 \end{tabular}
\caption[Table 1]{Comparing the power of the test statistics in  \eqref{DQ} for each $F_m$, $m=1,\dots,3$, in \eqref{fs} with that of the  K-2 rotated statistics in \eqref{DF}.
The true model from which the data are generated is that in \eqref{p}. Each simulation involves 100,000 replicates and $100$ observations. The significance levels considered are $\alpha=0.001$ ($3.29\sigma$), $\alpha=0.05$ ($1.96\sigma$), and $\alpha=0.1$ ($1.64\sigma$).}
\label{Tab1}
\end{table*} 

 Notice that, in practice, $\bm{\beta}$ and $\bm{\theta}$ are unknown. Hence, in order to compute steps 1-4, one can proceed by simply  plugging-in their MLEs $\widehat{\bm{\beta}}_{obs}$ and $\widehat{\bm{\theta}}_{obs}$ obtained on the observed data. In the case where $P\equiv F_{\bm{\beta}}$, $\widehat{\bm{\beta}}_{obs}$ converges, in probability, to the true value of $\bm{\beta}$, whereas, $\widehat{\bm{\theta}}_{obs}$ coverges to the values of $\bm{\theta}$ which minimizes the Kullback–Leibler divergence between $F_{\bm{\beta}}$ and $Q_{\bm{\theta}}$ \cite[e.g.,][p.147]{davison}.
The integrals can be computed as  Darboux sums over a grid of possible $\bm{x}$ values  on the search region $\mathcal{X}$. Finally, it is worth poiting out that  all the operators considered are linear, and thus, when $p$ is large, their implementation may be tedious but yet relatively simple; especially since they only need to be computed once in order to evaluate \eqref{DF} on the data observed.


\subsection{Empirical studies} To assess the performance of the testing procedure described above, we consider a dataset of $n=100$ observations generated from a bivariate Cauchy distribution, $P$, truncated over the range $\mathcal{X}=[1,20]\times[1,25]$, and  density
\begin{equation}
\label{p}
p(\bm{x})\propto (2\pi)^{-1}|\bm{\Sigma}|^{-1/2}\bigl[1+(\bm{x}-\mu)^{T}\bm{\Sigma}^{-1}(\bm{x}-\mu)\bigl]^{-3/2}
\end{equation}
where $\mu=(0,3)^{T}$, $\bm{\Sigma}$ is a matrix of diagonal elements $\sigma_{11}=\sigma_{22}=20$ and off-diagonal elements $\sigma_{12}=\sigma_{21}=10$.
Our goal is to test the validity of three different models for our data. Specifically, 
\begin{equation}
\begin{split}
\label{fs}
f_1(\bm{x};\bm{\beta})&\propto x_1^{(\beta_1-1)}x_2^{(\beta_2-1)}\exp\bigl\{-\beta_3(x_1+x_2)\bigl\},\\
f_2(\bm{x};\bm{\beta})&\propto\frac{\beta_3}{2\pi}[(x_1-\beta_1)^2+(x_2-\beta_2)^2+\beta_3]^{-3/2},\\
f_3(\bm{x};\bm{\beta})&\propto e^{-\frac{1}{200}\Bigl[\bigl(\frac{x_1}{\beta_1}-1\bigl)^2+\bigl(\frac{x_2}{\beta_2}-1\bigl)^2-\beta_3\bigl(\frac{x_1}{\beta_1}-1\bigl)\bigl(\frac{x_2}{\beta_2}-1\bigl)\Bigl]},
\end{split}
\end{equation}
that is, $f_1$ is the pdf of  a bivariate Gamma with independent components, $f_2$ is the pdf of a bivariate Cauchy with dependent component (but with dependence structure different from \eqref{p}), and $f_3$ is the pdf of a multivariate normal with dependent components. We denote with $F_1,F_2$ and $F_3$ the respective cdfs. Finally, we consider as reference distribution, $Q$,  the bivariate normal with independent components introduced in Section \ref{testingQ} and with pdf as in equation \eqref{q}.
Notice that all the models in \eqref{fs} are quite different from each other as well as from \eqref{q}. Moreover, each of these models is characterized by $p=3$ unknown parameters.

We proceed by simulating the null distributions of the three test statistics in \eqref{DQ} under $Q$ and their counterparts for each of the $F_m$, $m=1,2,3$, models considered; we denote the latter with 
 $\widehat{D}_{F_m}, \widehat{\omega}^2_{F_m}$ and $\widehat{A}^2_{F_m}$.  
The results are shown in the upper panels of Figure \ref{Fig2}. Despite the null distribution of the three statistics under $Q$ and $F_1$ appear fairly close, as expected, they are substantially different from those of $F_2$ and $F_3$. Therefore, in order to achieve distribution-freeness, we consider the test statistics in \eqref{DF} obtained by implementing Steps 1-4   in Section \ref{FtoQ}. We simulate their null distributions and we compare them with those of $\widehat{D}_Q, \widehat{\omega}^2_Q$, and $A^2_Q$, under model $Q$. The results are shown in bottom panels Figure \ref{Fig2}.

The  distributions of the K-2  rotated statistics $\widetilde{D}_{F_m}, \widetilde{\omega}^2_{F_m}$ and $\widetilde{A}^2_{F_m}$,  $m=1,2,3$, cannot be distinguished from those of $\widehat{D}_Q, \widehat{\omega}^2 _{Q}$ and $\widehat{A}^2 _{Q}$. Therefore, one can test $Q,F_1,F_2$ and $F_3$ by relying \underline{solely} on the simulated distribution of $\widehat{D}_Q, \widehat{\omega}^2 _{Q}$ and $\widehat{A}^2 _{Q}$, reducing the computational time by a factor of at least three (as we need to perform just one simulation instead of four).

Table \ref{Tab1} collects the results of a power study. There, we compare the power of the K-2 rotated test statistics in \eqref{DF}  with that of their classical counterparts   in \eqref{DQ}, and for different significance levels.  Interestingly, for model $F_2$, that is, the closest to the true distribution $P$ among those considered, the power of the K-2  rotated Kolmogorov-Smirnov and Cramer-von Mises statistics is higher compared to that of their non-rotated version. When testing $F_1$ and $F_3$, the power decreases for Kolmogorov-Smirnov. The power is comparably high in all the other cases. Notice that the power of the K-2 rotated statistics is not universally higher than their non-rotated counterparts. That is because, the K-2 rotated test statistics are simply new test statistics which may perform better than the classical Kolmogorov-Smirnov, Cramer-von Mises and Anderson Darling in some scenarios, but not in others.  
\section{Final remarks}
\label{Final}The K-2 transformation is a very powerful tool to achieve distribution-freeness in a simulation-based settings. Researchers can rely on simulations under a simplified model, $Q$, whose likelihood is easily accessible, and then construct suitable test statistics for one or more complex models $F$ which can be compared with the same simulated distribution.  

It is worth emphasizing that  the approximation of the null distribution of the statistics in \eqref{DF} with those of \eqref{DQ} does depend on the sample size. That is because the K-2 transform maps the limiting distribution of the process $\widetilde{v}_{F,n}(\bm{x},{\bm{\lambda}})$ into that of $\widetilde{v}_{Q,n}(\bm{x},{\bm{\theta}})$. In light of this,  in order to achieve a good approximation for moderately large samples (e.g., 100 observations),   it is recommendend to choose $Q$ ``sufficiently close to $F$'' so that the entire search region is sampled reasonably often under both $Q$ and  $F$.

To compute the K-2 rotation, one needs to evaluate the score functions of  $F$. In situations where the likelihood is not tractable in closed-form, a possible solution is that of constructing templates for the score, starting from the likelihood templates and applying the definition of derivative. Their evaluation does not need to be repeated on multiple runs, and it is only needed to evaluate the K-2 rotated test statistics on the data observed.

\acknowledgments
The author  thanks an anonymous referee whose feedback
has been substantial to improve the overall clarity of the
paper.

\appendix
\section{\textbf{Deriving equation \eqref{projection1}}}
\label{app}
Consider the empirical process
\begin{equation}
\label{vhat}
\vQn(\bm{x},\widehat{\bm{\theta}})=\frac{1}{\sqrt{n}}\sum_{i=1}^n\psi_{\bm{x},\widehat{\bm{\theta}}}(\bm{x}_i).
\end{equation}
and the vectors of derivatives
$\bm{\dot{\psi}}_{\bm{x},\bm{\theta}}(\bm{t})$ and $\bm{\dot{q}}_{\bm{\theta}}(\bm{t})$ with components 
\begin{align}
\dot{\psi}_{\bm{x},\bm{\theta}_j}(\bm{t})&=\frac{d}{d\theta_j}\psi_x(\bm{t})\quad \text{and}\\
\dot{q}_{\bm{\theta}_j}(\bm{t})&=\frac{d}{d\theta_j}q_{\bm{\theta}}(\bm{t}).
\end{align}
Where,
 \begin{align}
 \label{a0000}
\dot{\psi}_{\bm{x},\bm{\theta}_j}(\bm{t})&= \frac{d}{d\widehat{\theta}_j}\psi_{x\bm{\theta}}(\bm{t})|_{\widehat{\bm{\theta}}=\bm{\theta}}=- \frac{d}{d{\theta_j}}Q_{\bm{\theta}}(\bm{x})\\
\label{a000}
&=- \frac{d}{d{\theta_j}}\int_{-\infty}^{\bm{x}}   q_{\bm{\theta}}(\bm{t})\text{d}\bm{t}=-\int_{-\infty}^{\bm{x}}  \dot{q}_{\bm{\theta}_j}(\bm{t})\text{d}\bm{t}\\
\label{a00}
&=-\int_{-\infty}^{\bm{x}}   \frac{\dot{q}_{\bm{\theta}_j}(\bm{t})}{{q}_{\bm{\theta}}(\bm{t})}q_{\bm{\theta}}(\bm{t})\text{d}\bm{t}.
 \end{align}
 where the integrals in \eqref{a0000}-\eqref{a00} are all multidimensional.
A Taylor expansion of \eqref{vhat} leads to

{\fontsize{9.5}{9.5}\selectfont{
\begin{align}
\label{a0}
\vQn(\bm{x},\widehat{\bm{\theta}})&\approx\frac{1}{\sqrt{n}}\sum_{i=1}^n\psi_{\bm{x},\bm{\theta}}(\bm{x}_i)+(\widehat{\bm{\theta}}-\bm{\theta})^{T}\frac{1}{\sqrt{n}}\sum_{i=1}^n\dot{\psi}_{\bm{x},\bm{\theta}}(\bm{x}_i)
\end{align}}}

The asymptotic expansion of $(\widehat{\bm{\theta}}-\bm{\theta})$ \citep[e.g.,][p. 53]{van} is
\begin{align}
\label{asy}
\sqrt{n}(\widehat{\bm{\theta}}-\bm{\theta})&=\frac{1}{\sqrt{n}}\Gamma_{\bm{\theta}}^{-1}\sum_{i=1}^n\frac{\dot{\bm{q}}_{\bm{\theta}}(\bm{x}_i)}{q_{\bm{\theta}}(\bm{x}_i)} +o_p(1)\\
\label{scoreb1Da}
&=\frac{1}{\sqrt{n}}\Gamma_{\bm{\theta}}^{-1/2}\sum_{i=1}^n\Gamma_{\bm{\theta}}^{-1/2}\bm{u}_{\bm{\theta}}(\bm{x}_i)+o_p(1)\\
\label{scoreb1D}
&=\frac{1}{\sqrt{n}}\Gamma_{\bm{\theta}}^{-1/2}\sum_{i=1}^n\bm{b}_{\bm{\theta}}(\bm{x}_i)+o_p(1)
\end{align}
where, as in \eqref{scoreb}, $\Gamma_{\bm{\theta}}$ is the Fisher information matrix, and $\bm{b}_{\bm{\theta}}(\bm{x})$
is vector of normalized score functions $b_{\bm{\theta}_j}(\bm{x})$.
Combining \eqref{a00}, \eqref{a0}, \eqref{asy} and \eqref{scoreb1D} we have
\begin{align}
\label{a2}
\vQn(\bm{x},\widehat{\theta})&\approx\frac{1}{\sqrt{n}}\sum_{i=1}^n\psi_{\bm{x},\bm{\theta}}(\bm{x}_i)\\
&-\frac{1}{\sqrt{n}}\sum_{i=1}^n\bm{b}_{\bm{\theta}}^{T}(\bm{x}_i) \int_{-\infty}^{\bm{x}}  \bm{b}_{\bm{\theta}}(\bm{t}) q_{\bm{\theta}}(\bm{t})\text{d}\bm{t}
\end{align}
where the error of the approximation  has be shown by \citet{khm80} to be $o_p(1)$.
Moreover, simple algebra can be applied to show that 
$<b_{\bm{\theta}_j},\psi_{\bm{x},\bm{\theta}}>_Q=\int_{-\infty}^{\bm{x}}  b_{\bm{\theta}_j}({\bm{t}})q_{\bm{\theta}}({\bm{t}})\text{d}{\bm{t}}$.
Specifically,
\begin{align}
\label{a4}
<&b_{\bm{\theta}_j},\psi_{\bm{x},\bm{\theta}}>_Q=\int_{-\infty}^\infty  b_{\bm{\theta}_j}(\bm{t})\psi_{\bm{x},\bm{\theta}}(\bm{t})q_{\bm{\theta}}(\bm{t})\text{d}\bm{t}\\
\label{a5}
&=\int_{-\infty}^\infty  b_{\bm{\theta}_j}(\bm{t})[\mathbbm{1}_{\{\bm{t}\leq \bm{x}\}}-Q_{\bm{\theta}}(\bm{x})]q_{\bm{\theta}}(\bm{t})\text{d}\bm{t}\\
\label{a6}
&=\int_{-\infty}^{\bm{x}}  b_{\bm{\theta}_j}(\bm{t})q_{\bm{\theta}}(\bm{t})\text{d}\bm{t}-Q_{\bm{\theta}}(\bm{x})\int_{-\infty}^\infty b_{\bm{\theta}_j}(\bm{t})q_{\bm{\theta}}(\bm{t})\text{d}\bm{t}\\
\label{a7}
&=\int_{-\infty}^{\bm{x}}  b_{\bm{\theta}_j}(\bm{t})q_{\bm{\theta}}(\bm{t})\text{d}\bm{t}
\end{align}
where \eqref{a7} follows from \eqref{a6}, and the fact that the normalized score vector $\bm{b}_{\bm{\theta}}$ has mean zero under $Q_{\bm{\theta}}$.
Finally, combining \eqref{a2} and \eqref{a4}-\eqref{a7}, we obtain
\begin{align}
\label{final}
\vQn(\bm{x},\widehat{\theta})&=\frac{1}{\sqrt{n}}\sum_{i=1}^n\widetilde{\psi}_{\bm{x},\bm{\theta}}({\bm{x}}_i)+o_p(1)
\end{align}
where $\widetilde{\psi}_{\bm{x},\bm{\theta}}({\bm{x}}_i)=\psi_{\bm{x},\bm{\theta}}(\bm{x}_i)-\sum_{j=1}^pb_{\bm{\theta}_j}(\bm{x}_i)<b_{\bm{\theta}_j},\psi_{\bm{x},\bm{\theta}}>_{Q_{\bm{\theta}}} $.

\bibliography{biblioPoi2}

\end{document}